\UseRawInputEncoding

\documentclass[english]{cccconf}
\usepackage[comma,numbers,square,sort&compress]{natbib}
\usepackage{epstopdf}

\usepackage{booktabs}

\usepackage{mathrsfs}

\usepackage{amsmath}

\usepackage{amssymb}

\usepackage{gensymb}  

\usepackage{appendix}

\usepackage{caption}

\usepackage{graphicx, subfig}

\graphicspath{{picures/}}

\begin{document}

\title{Quantum Graph Convolutional Neural Networks}

\author{Jin Zheng\aref{amss},
	Qing Gao\aref{amss,hit,}$^{*}$,           
	Yanxuan L\"u\aref{amss}
}

\affiliation[amss]{School of Automation Science and Electrical Engineering, Beihang University, Beijing 100191, P.~R.~China}

\affiliation[hit]{Beijing Advanced Innovation Center for Big Data and Brain Computing, Beihang University, Beijing 100191, P.~R.~China  \email{zhengjin03@buaa.edu.cn; gaoqing@buaa.edu.cn; lvyanxuan@buaa.edu.cn}
}

\maketitle

\begin{abstract}
At present, there are a large number of quantum neural network models to deal with Euclidean spatial data, while little research have been conducted on non-Euclidean spatial data. In this paper, we propose a novel quantum graph convolutional neural network (QGCN) model based on quantum parametric circuits and utilize the computing power of quantum systems to accomplish graph classification tasks in traditional machine learning. The proposed QGCN model has a similar architecture as the classical graph convolutional neural networks, which can illustrate the topology of the graph type data and efficiently learn the hidden layer representation of node features as well. Numerical simulation results on a graph dataset demonstrate that the proposed model can be effectively trained and has good performance in graph level classification tasks.
\end{abstract}


\footnotetext{This work is supported by National Natural Science
	Foundation of China (No. 61903016, No. 61803132) and by Alexander von Humboldt Fellowship, Germany.}

\section{Introduction}

Machine learning and deep learning have become powerful tools for mining data patterns and technical cornerstones in the era of big data. With continuous development of information technology, new data types, structures, and production speeds have exploded, which causes machine learning to encounter computational bottlenecks when processing high-dimensional data. As another important research frontier, quantum computing has attracted tremendous attention from the artificial intelligence community. The unique characteristics of superposition, entanglement and quantum measurement make quantum systems extraordinary computing medium, especially when processing high-dimensional data, that is, quantum computing shows potential exponential acceleration for machine learning compared with classical computing. As an emerging interdisciplinary field from both quantum computing and machine learnning, quantum machine learning is dedicated to taking the advantages of quantum computing to improve classical machine learning algorithms \cite{Biamonte2017,wurebing2017}. Although there are significant challenges in hardware and software implementation, quantum machine learning has become a powerful machine learning application.

In quantum machine learning, quantum neural networks combine the advantages of quantum computing and neural network models \cite{Gao2020}. They have the potential to improve the computational efficiency of classical neural networks significantly. KAK first proposed a quantum neural network in 1994 \cite{Kak1995}. Then scholars extensively explored a variety of possible quantum neural network models based on the noisy mesoscale quantum devices \cite{wurebing2020}, such as the quantum perceptron model \cite{Kapoor2016}, the quantum tensor neural network \cite{Schütt2017}, the quantum convolutional neural networks \cite{Cong2019}, etc. These quantum neural network models simulate typical quantum systems with network structural characteristics in the quantum Hilbert space and address the higher complexity and dimensionality issues in classical neural networks.

Despite the aforementioned significant progress that have been achieved, most existing quantum neural networks are designed to deal with regularly structured data in European space. The tensor calculation system based on these data is relatively natural and efficient. For data in non-Euclidean space, classical machine learning algorithms use graph neural networks to structure the learning process of graph data directly \cite{Kipf2016}. Nevertheless, there is few corresponding research in the quantum field. In this paper, we propose a novel quantum graph convolutional neural network model implementable on quantum parameterized circuits. The model is motivated by the quantum convolutional neural networks \cite{Cong2019}, classical graph convolutional neural networks \cite{Kipf2016}, and the quantum advantages with shallow circuits \cite{Bravyi2018}. The model design starts from the overall structure of the graph and aims to solve a classification task at the graph level. The design procedure can be summarized as follows:

Step 1: The graph data are effectively encoded into quantum states via the amplitude encoding method.

Step 2: Based on the graph structure, a series of parameter-based universal quantum gates are used to construct the quantum graph convolutional neural network after the amplitude encoding block.

Step 3: The learning ability of the quantum graph convolutional neural network model is tested by training on a quantum circuit using a specific dataset.

The rest of this article is organized as follows: the second section describes the prerequisite knowledge of quantum computing and graph convolutional neural networks. The third section introduces the concrete realization scheme of the quantum graph convolutional neural network model. The fourth section provides experimental simulation and results analysis. The fifth section discusses experimental conclusions and future work.

\section{Preliminaries}

To interpret the quantum graph convolutional neural network model some basic pre-knowledge of quantum mechanics and graph neural networks theory is essential.

\subsection{Quantum Computing}

\subsubsection{Quantum States}

Different from classical computers that calculate using bits, the basic computational unit of a quantum computer is the quantum bit, or the qubit. A qubit has two basis states, i.e., $\left| 0 \right\rangle$ and $\left| 1 \right\rangle$, corresponding to the ground state and the excited state of a two-level quantum system. A pure quantum state in the same Hilbert space can be always written as 
$\left| \psi  \right\rangle =\alpha \left| 0 \right\rangle+\beta \left| 1 \right\rangle$, 
where $\alpha ,\beta$ are complex numbers
satisfying 
${{\left| \alpha  \right|}^{2}}+{{\left| \beta  \right|}^{2}}=1$, 
or 
 $\left| \psi  \right\rangle ={{e}^{i\gamma }}\left( \cos \left( \frac{1}{2}\theta  \right)\left| 0 \right\rangle +{{e}^{i\phi }}\sin \left( \frac{1}{2}\theta  \right)\left| 1 \right\rangle  \right)$, 
where 
$\theta$, 
$\phi$ and $\gamma$ are real numbers with 
$0\le \theta \le \pi $ 
and $0\le \phi <2\pi $. 
The $n$-qubit space is formed by the tensor product of $n$ single-qubit Hilbert spaces. 
For a vector 
$x\in {\mathbb{R}^{{{2}^{n}}}}$, 
the amplitude encoded \cite{Zhang2020} state $\left| x \right\rangle$ 
is defined as $\frac{1}{\left\| x \right\|}\sum\nolimits_{i=1}^{{{2}^{n}}}{{{x}_{i}}\left| i \right\rangle }$.
Note that the amplitude of a quantum state is not always real. Instead, in most cases, they are complex numbers. 

\subsubsection{Quantum Gates and Quantum Circuits}

In quantum circuits, a single-qubit operation to the state behaves like the matrix-vector multiplication and can be referred as the gate.

(1) single-qubit gates

The single-qubit gates used in the model are Pauli-rotation gates. They are generated by taking exponentials of the Pauli operators (including ${{R}_{x}}$, ${{R}_{y}}$, and ${{R}_{z}}$), which can rotate the state vector by an arbitrary angle about the corresponding axis of Bloch sphere, as explained in Table~\ref{tab1}.

\begin{table}[!hb]
	\centering
	\caption{Pauli-rotation gates}
	\label{tab1}
	\begin{tabular}{l|l}
		\hhline
		Pauli-rotation gates & Matrix representations \\ \hline
		${{R}_{x}}$ gate & 
		${{R}_{x}}\left( \theta  \right)$=
		$\begin{bmatrix}
			\cos \left( \frac{1}{2}\theta  \right) & -i\sin \left( \frac{1}{2}\theta  \right) \\
			-i\sin \left( \frac{1}{2}\theta  \right) & \cos \left( \frac{1}{2}\theta  \right)
		\end{bmatrix}$ \\ \hline 
		${{R}_{y}}$ gate & 
		${{R}_{y}}\left( \theta  \right)$=
		$\begin{bmatrix}
			\cos \left( \frac{1}{2}\theta  \right) & -\sin \left( \frac{1}{2}\theta  \right) \\
			\sin \left( \frac{1}{2}\theta  \right) & \cos \left( \frac{1}{2}\theta  \right)
		\end{bmatrix}$ \\ \hline
		${{R}_{z}}$ gate & 
		${{R}_{z}}\left( \theta  \right)$=
		$\begin{bmatrix}
			{{e}^{-i\frac{1}{2}\theta }} & 0 \\
			0 & {{e}^{+i\frac{1}{2}\theta }}
		\end{bmatrix}$ \\
		\hhline
	\end{tabular}
\end{table}

(2)	two-qubit gates

The two-qubit gates required in the quantum graph convolutional neural network proposed in the paper are the CNOT gate and the SWAP gate. The CNOT gate is important to operate on more qubits at the same time. The matrix representations of a CNOT gate and a SWAP gate are shown in Table~\ref{tab2}.

\begin{table}[!htb]
	\centering
	\caption{2-qubit gates}
	\label{tab2}
	\begin{tabular}{l|l}
		\hhline
		2-qubit gates          & Matrix representations \\ \hline
		CNOT gate & 
		$\begin{bmatrix}
			1 & 0 & 0 & 0 \\
			0 & 1 & 0 & 0 \\
			0 & 0 & 0 & 1 \\
			0 & 0 & 1 & 0 
		\end{bmatrix}$ \\ \hline
		SWAP gate & 
		$\begin{bmatrix}
			1 & 0 & 0 & 0 \\
			0 & 0 & 1 & 0 \\
			0 & 1 & 0 & 0 \\
			0 & 0 & 0 & 1 	
	\end{bmatrix}$ \\ \hhline
	\end{tabular}
\end{table}

\subsubsection{Quantum Measurement}

After the data undergoes the unitary transformation of the quantum gate, the transformation result is not directly accessible, and we need to perform quantum measurements to obtain the results. For example, performing a projective measurement with $Z$ observable on the qubit with the state $\left| \psi \right\rangle \ =\alpha \left| 0 \right\rangle \ +\beta \left| 1 \right\rangle$ generates 1 and -1 with probability 
$p\left( 1 \right)={{\left| \alpha  \right|}^{2}}$ and 
$p\left( -1 \right)={{\left| \beta  \right|}^{2}}$, respectively. 
After the measurement, the qubit collapses to the new state $\left| 0 \right\rangle $ or $\left| 1 \right\rangle $.

The calculation basis measurement can be understood as extracting a sample of a binary string from the distribution defined by the quantum state \cite{Chung2000}. However, the measurement results are probabilistic, and one possible value can be obtained in each measurement under some probability. In order to obtain as accurate information about the quantum state as possible, we need to perform repeated measurements.

\subsection{Graph Convolutional Neural Networks}

Convolutional neural networks have local connection and weight sharing  properties \cite{Krizhevsky2017}, which correspond with the regular 2D grid structure of image data. A lot of outstanding works have been done on data research with a regular spatial structure. However, many kinds of data do not have regular spatial structures, such as the graph structure data. When either the number of neighbor nodes or nodes order is uncertain, it becomes challenging to select a fixed convolutional kernel to adapt to the entire graph's irregularity. The graph convolutional neural networks can automatically learn node characteristics and learn the associated information between nodes. Its core idea is to use edge information to aggregate node information to generate new node representations.

\begin{figure}[!htb]
	\centering
	\includegraphics[width=0.5\hsize]{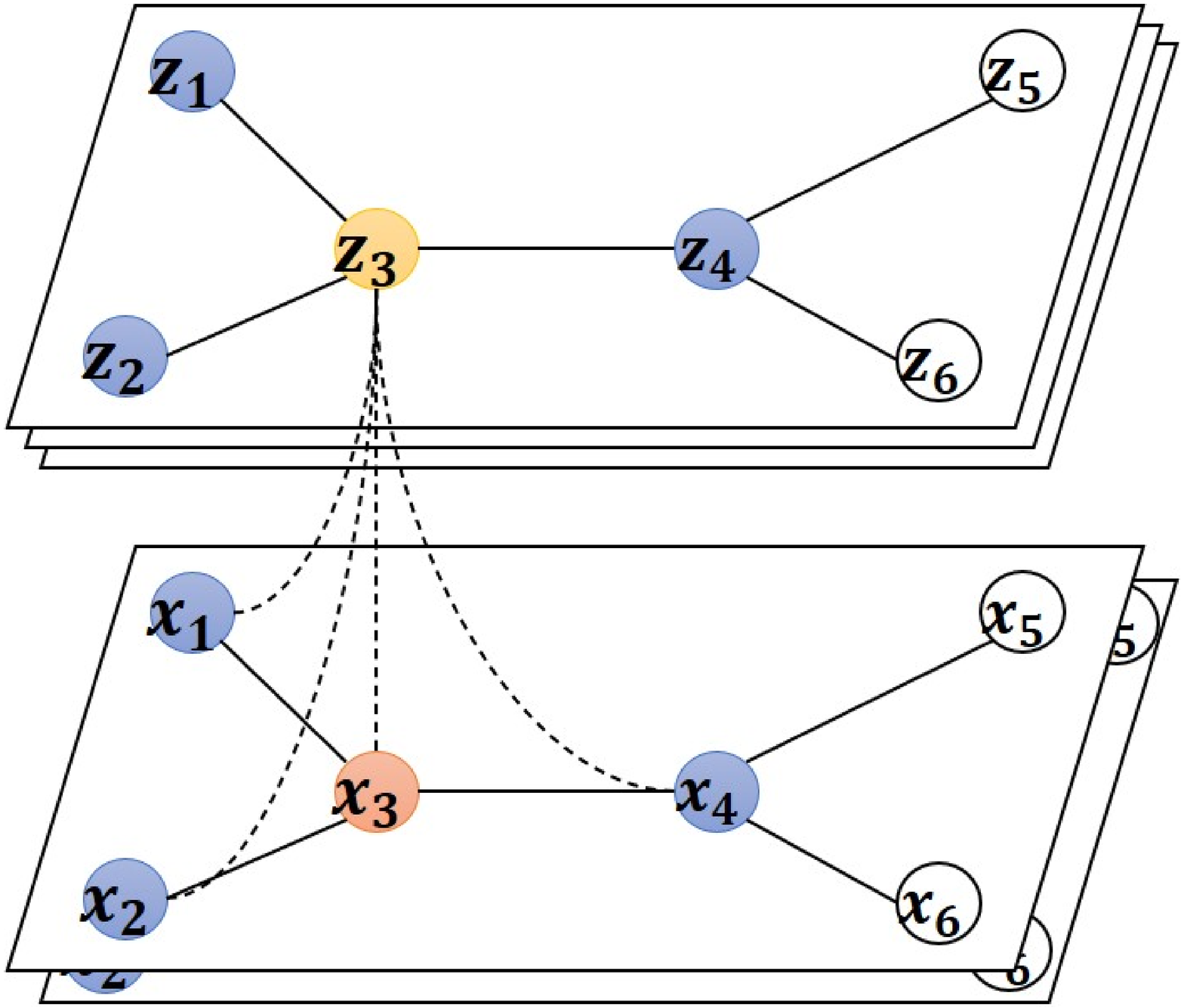}
	\caption{Graph convolutional neural networks}
	\label{fig1}
\end{figure}

Given a graph $G$, ${{x}_{v}}$ denotes the feature vector of a node $v$. The learning goal of the graph convolutional neural networks is to obtain the hidden state of each node, which contains information from the neighboring nodes. As in Fig.~\ref{fig1}, ${{x}_{3}}$ aggregates the features of its neighbor nodes ${{x}_{1}}$, ${{x}_{2}}$, and ${{x}_{4}}$ through the formula~(\ref{eq1}).

\begin{equation}
	\label{eq1}
	{{H}^{\left( l \right)}}\left( v \right)=\sigma \left( \sum\nolimits_{u\in N\left( v \right)}{{{H}^{\left( l \right)}}\left( u \right){{W}^{\left( l \right)}}} \right),
\end{equation} where ${H}^{\left( l \right)}$ is the hidden state of the node $v$ at time $l$, and $W$ is the feature dimension conversion matrix of the aggregate node features. The hidden state is updated iteratively according to (\ref{eq2}) at time $l+1$:
\begin{equation}
	\label{eq2}
	{{H}^{\left( l+1 \right)}}=\left( {{H}^{\left( l \right)}},A \right)=\sigma \left( A{{H}^{\left( l \right)}}{{W}^{\left( l \right)}} \right),
\end{equation} where $A$ represents the adjacency matrix, and  
$A{{H}^{\left( l \right)}}{{W}^{\left( l \right)}}$ 
selects the first-order neighbor nodes to realize the information transmission. However, the formula~(\ref{eq2}) does not consider the information of the node itself during aggregation. By adding a closed loop to the adjacency matrix and normalizing it, the expression of the graph convolutional neural networks can be obtained \cite{Defferrard2016} as:
\begin{equation}
	\label{eq3}
	H^{(l+1)} = f(H^l,A)= \sigma({\tilde{D}}^{-\tfrac{1} {2}}\tilde{A}\tilde{D}^{-\tfrac{1} {2}}H^{(l)}W^{(l)}).
\end{equation}

\section{A Quantum Graph Convolutional Neural Network Model}

\begin{figure}[!htb]
	\centering
	\includegraphics[width=0.8\hsize]{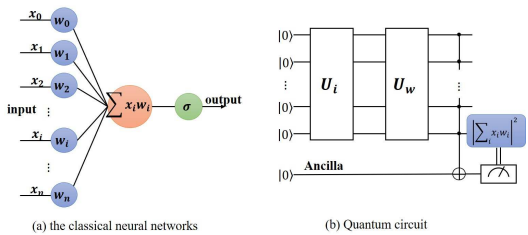}
	\caption{The classical neural network and a typical quantum circuit}
	\label{fig2}
\end{figure}

The recently proposed quantum neural networks refer to computational models with networked structures and trainable parameters realized by quantum circuits \cite{Killoran2019}. In 2018, Tacchino implemented the world's first single-layer neural network \cite{Tacchino2019} on a quantum computer. The classical neural network with a single neuron is shown in Fig.~\ref{fig2}(a). The input is weighted and summed, and is then mapped to the output through the activation function. The quantum neural networks have the same process. However, the implementation of the quantum processors is different from the classical computers, as illustrated in Fig.~\ref{fig2}(b). The first layer of quantum neural networks encodes the input data into quantum states. The second layer performs a unitary transformation on the input, similar to the weight vectors in the classical neural networks. Finally, the output is written into the ancilla qubit, and the final result is generated via quantum measurements.

Combing the characteristics of the graph type data and the parameterized quantum circuits, in this paper, we propose a novel quantum graph convolutional neural network model that consists of four blocks: quantum state preparation, quantum graph convolutional layers, quantum pooling layers, and quantum measurements. The model aggregates the node information based on the relationship between edges and using machine learning algorithms to accomplish the graph classification tasks.

\begin{figure}[!htb]
	\centering
	\includegraphics[width=0.3\hsize]{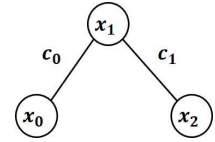}
	\caption{A simple graph data structure example}
	\label{fig3}
\end{figure}

\begin{figure}[!htb]
	\centering
	\includegraphics[width=0.8\hsize]{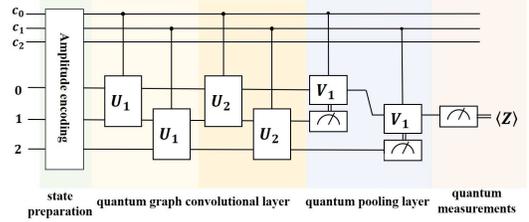}
	\caption{A QGCN quantum circuit model}
	\label{fig4}
\end{figure}

Here we demonstrate the overall framework of QGCN in details. The classical graph data is encoded into quantum states that can be processed by quantum circuits. The model consists of a series of parameter-based unitary transformations and extracts the features in quantum states. The quantum measurements block outputs the expectation values $\left\langle Z \right\rangle$ as the classification results. The universal quantum graph convolutional neural networks are composed of these four blocks. The QGCN model can be expanded according to different graph data structures. For example, the graph data structure is Fig.~\ref{fig3}, its corresponding QGCN quantum circuit model is Fig.~\ref{fig4}.

\subsection{Quantum State Preparation}

An efficient procedure to encode classical data in quantum superposition states can genuinely realize the advantages of quantum computing, so state preparation is an essential block in the quantum graph convolutional neural networks. Quantum machine learning algorithm usually uses amplitude encoding as quantum state preparation because its coding property is similar to the original vector. A normalized classical vector $x\in {{\mathbb{C}}^{{{2}^{n}}}}$, $\sum\nolimits_{j}{{{\left| {{x}_{j}} \right|}^{2}}=1}$ can be represented by a quantum state $\left| \psi  \right\rangle \in \mathcal{H}$ as follows:
\begin{equation}
	\label{eq4}
	x=\left( \begin{matrix}
		{{x}_{1}}  \\
		\vdots   \\
		{{x}_{{{2}^{n}}}}  \\
	\end{matrix} \right)\leftrightarrow \left| {{\psi }_{x}} \right\rangle =\sum\nolimits_{j=1}^{{{2}^{n}}}{{{x}_{j}}\left| j \right\rangle }.
\end{equation}
In the same fashion, a classical matrix $B\in {{\mathbb{C}}^{{{2}^{n}}\times {{2}^{m}}}}$ with entries ${{a}_{ij}}$ that satisfies $\sum\nolimits_{ij}{{{\left| {{a}_{ij}} \right|}^{2}}=1}$, can be encoded as $\left| {{\psi }_{B}} \right\rangle =\sum\limits_{i=1}^{{{2}^{m}}}{\sum\limits_{j=1}^{{{2}^{n}}}{{{a}_{ij}}\left| i \right\rangle \left| j \right\rangle }}$ by enlarging the Hilbert space accordingly.

Note that amplitude encoding can only process normalized classical vectors, i.e., $\left| x \right\rangle =\sum\nolimits_{j=1}^{{{2}^{n}}}{\frac{{{x}_{j}}}{\left\| x \right\|}\left| j \right\rangle }$. A classical 2-dimensional vector $\left( x,y \right)$ can only be associated with an amplitude vector $\left( \alpha ,\beta  \right)$ of a qubit, which satisfies ${{\left| \alpha  \right|}^{2}}\text{+}{{\left| \beta  \right|}^{2}}\text{=}1$. To solve this problem effectively, we can increase the classical vector space by one dimension and normalize the resulting vector. Then, the $N$-dimensional space will be embedded in $N{+}1$-dimensional space in which the data is normalized without loss of information \cite{Park2019}. For example, encoding the vector $x=\left( 0.1,-0.6,1.0 \right)$ by amplitude encoding, we have to first normalize it to unit length and pad it with 0 to a dimension of integer logarithm, ${{x}^{'}}=\left( 0.073,-0.438,0.730,0.000 \right)$. Then, ${{x}^{'}}$ can be represented by a quantum state of 2 qubits, $0.073\left| 00 \right\rangle -0.438\left| 01 \right\rangle +0.730\left| 10 \right\rangle +0\left| 11 \right\rangle $. 
Similarly, the state encodes the matrix 
$B=\left( \begin{matrix}
	0.073 & -0.438  \\
	0.730 & 0.000  \\
\end{matrix} \right)$.

\subsection{Quantum Graph Convolutional Layer}

Similar to the function of the classical graph convolutional layer, the quantum graph convolutional layer can also realize local connection and parameter sharing properties by implementing two-qubit unitary operations $U$. 

We use two sets of qubits to describe the nodes and the topology of a graph dataset, respectively. To be specific, a number of qubits in either the excited state $\left| 1 \right\rangle$ or the ground state $\left| 0 \right\rangle$ is used to denote the local connectivity information, that is, if two nodes are connected, the corresponding qubit is set to be $\left| 1 \right\rangle$ and affects the two nodes through a controlled unitary gate. On the other hand, a number of qubits are used to encode the feature information of the nodes through the amplitude encoding method. In this way, the whole architecture of the QGCN model can adapt to different kinds of graph type datasets, which uses the topology (or the adjacent matrix) and the node features as the inputs. When facing node level tasks, each single sample from the graph type dataset must provide both the adjacent matrix and the node feature information; when facing graph level tasks, the QGCN would have a fixed and generally simpler structure and only the node features need to be considered as the input.

The quantum graph convolutional layer represents the aggregation of neighbor nodes to the current node. The number of layers means the order of node aggregation, so the unitary operations in the same layer have the same parameters, reflecting parameter sharing characteristic.

\begin{figure}[!htb]
	\centering
	\includegraphics[width=0.5\hsize]{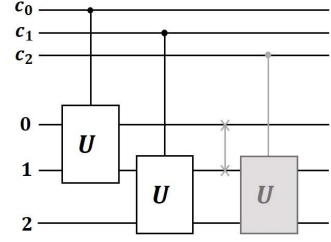}
	\caption{A quantum graph convolutional layer}
	\label{fig5}
\end{figure}

In Fig.~\ref{fig5}, we provide a quantum circuit corresponding to the quantum graph convolutional layer of Fig.~\ref{fig3}. The depth of the graph is two. Since each quantum graph convolutional layer implements unitary operations with different parameters but the same structure, Fig.~\ref{fig5} only shows the circuit of one particular layer. ${x_0}$, ${x_1}$ and ${x_2}$ aggregate its first-order neighbor nodes respectively. There is no edge connection between ${x_0}$ and ${x_2}$, so ${c_2}$ does not work, and unitary operations are not performed.

Since the quantum graph convolutional layer can be expanded based on the graph data structures, the designed model should be easy to encapsulate and have generalization capabilities. Therefore, it is necessary to implement an arbitrary unitary two-qubit operation with a minimal number of gates. The minimum implementations for two-qubit gates, \cite{Shende2004} specified the minimum number of gates for an arbitrary unitary matrix $U\in SU\left( 4 \right)$ is 18, including 3 CNOT gates, 3 rotation gates, and 4 arbitrary single-qubit gates. For instance, Vidal and Dawson proved that any two-qubit unitary operations could be expressed with three or fewer CONT gates and some one-qubit rotations \cite{Vidal2004}. For arbitrary single-qubit gates, they are decomposed into the rotation gates, as in the formula~(\ref{eq5}):
\begin{align}
	\label{eq5}
	{{U}_{1}} = 
	\left(\begin{smallmatrix}
		e^{i\delta} & 0 \\
		0 & e^{i\delta}
	\end{smallmatrix} \right)
	\left(\begin{smallmatrix}
		e^{\frac{i\alpha} {2}} & 0 \\
		0 & e^{-\frac{i\alpha} {2}}
	\end{smallmatrix} \right)
	\left(\begin{smallmatrix}
		cos(\frac{\gamma} {2}) & sin(\frac{\gamma} {2}) \\
		sin(-\frac{\gamma} {2}) & cos(\frac{\gamma} {2})
	\end{smallmatrix} \right)
	\left(\begin{smallmatrix}
		e^{\frac{i\beta} {2}} & 0 \\
		0 & e^{-\frac{i\beta} {2}}
	\end{smallmatrix} \right),
\end{align}
where $\delta ,\alpha ,\beta $ and $\gamma $ are real angles. This corresponds to two ratation-$z$ gates, and one rotation-$y$ gate. For any unitary matrix ${{U}_{1}}\in SU\left( 2 \right)$, the angle $\Phi ,\alpha ,\beta ,\gamma $ can be found so that the following formula~(\ref{eq6}), formula~(\ref{eq7}), formula~(\ref{eq8}) are satisfied:
\begin{equation}
	\label{eq6}
	{{U}_{1}}={{e}^{-i\Phi }}*
	\begin{pmatrix}
		{A} & {B}  \\
		{C} & {D}  
	\end{pmatrix}, 
\end{equation}

\begin{equation}
	\label{eq7}
	{{U}_{1}}={{e}^{-i\Phi }}{{R}_{z}}\left( \alpha  \right){{R}_{y}}\left( \beta  \right){{R}_{z}}\left( \gamma  \right),
\end{equation}

\begin{equation}
	\label{eq8}
	S{{U}_{1}}\left( 2 \right)={{R}_{z}}\left( \alpha  \right){{R}_{y}}\left( \beta  \right){{R}_{z}}\left( \gamma  \right).
\end{equation}

\begin{figure}[!htb]
	\centering
	\includegraphics[width=0.8\hsize]{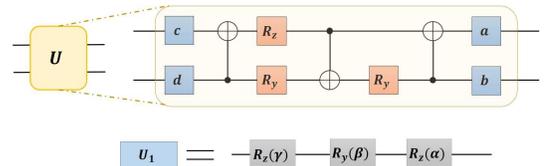}
	\caption{The universal two-qubit gate circuit}
	\label{fig6}
\end{figure}

In this paper, we apply a universal two-qubit gate circuit composed of 3 CNOT gates and 15 basic gates from $\left\{ CNOT,{{R}_{y}},{{R}_{z}} \right\}$ to implement the convolutional operation, as in Fig.~\ref{fig6}. The quantum gates ${a,b,c,d}\in SU\left( 2 \right)$ are the universal single-qubit gates.

\subsection{Quantum Pooling Layer}

\begin{figure}[!htb]
	\centering
	\includegraphics[width=0.8\hsize]{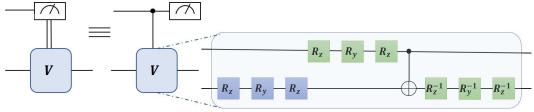}
	\caption{A quantum pooling layer circuit}
	\label{fig7}
\end{figure}

The model of the quantum pooling layer is different from the quantum graph convolutional layer. It reduces the feature dimension by introducing quantum measurements and achieves the same effect as the classical pooling layer. The model first measures a portion of qubits and then determines whether to implement unitary transformations ${{V}_{i}}$ on the neighbor qubits based on the measurement results. To design the quantum pooling model more general, the gates should have an arbitrary control state and apply arbitrary single-qubit transformations. Therefore, the quantum circuit includes a CNOT gate, the parametrized ${{R}_{y}}$, ${{R}_{z}}$ gates, and inverse rotation gates $R_{y}^{-1}$, $R_{z}^{-1}$ as shown in Fig.~\ref{fig7}.

\subsection{Quantum Measurement}

After multi-layer convolutional and pooling transformations, the quantum measurement is performed on the specified qubit to obtain the expectation value. The $Z$ observable is used in the QGCN model, and the final discrimination result is obtained by mapping the expected value to the model output. For example, for the binary classification tasks of the graph, $\left\langle Z\right\rangle\ge0$ indicates dividing the sample to the positive class, and $\left\langle Z\right\rangle\ < 0$ indicates dividing the sample to the negative class.

\section{Experiments and Discussions}

In this section, the learning ability of the quantum graph convolutional neural network model is tested by training on the quantum circuit using a specific dataset. All experiments are simulated based on the Pennylane Python package \cite{Ville2020}. Pennylane provides a bridge between classical computing and quantum computing with a default simulator device. The simulator device is well integrated with external software and hardware and can simulate real quantum computers to run quantum circuits.

\subsection{Datasets Preprocessing}

\begin{figure}[!htb]
	\centering
	\includegraphics[width=0.8\hsize]{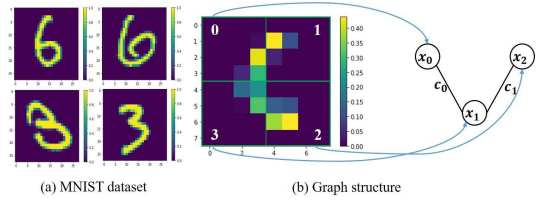}
	\caption{The experimental dataset}
	\label{fig8}
\end{figure}

The experiment uses the MNIST dataset \cite{LeCun1998} to construct the graph data structure. The original MNIST database of handwritten digits has 60,000 training samples and 10,000 test samples. Each sample includes a 28$\times $28 grayscale image of a handwritten number and its corresponding label. As the classification dataset, we randomly select 480 samples labeled 3 and 6 from the training set to build a small sample training dataset, and similarly build a test dataset with 120 samples, as shown in Fig.~\ref{fig8}(a). Then the classification data is downsampled as an 8$\times $8 grayscale image to construct a graph structure. Every 4$\times $4 grid can be used as a graph node. The position relationship between grids corresponds to the edge of nodes, and each node is numbered clockwise. The experiment randomly selects three nodes to construct the graph structure from each graph in the dataset. For example, if the numbers 0, 2, and 3 are chosen as the graph nodes, the corresponding graph data structure is as in Fig.~\ref{fig8}(b). Note that the dataset needs to be normalized to fit the amplitude encoding. The label of digit 3 is set to 1 and -1 for digit 6.

\subsection{Quantum State Preparation}

Amplitude encoding associates classical information with quantum amplitudes. The operation of encoding classical floating-point data into quantum probability amplitudes is strictly limited and can only process normalized classical vectors. The features of each node in the experimental data are encoded to quantum probability magnitudes, and then tensor product operations are performed. By expanding the Hilbert space and normalizing the original numerical vectors, all nodes and edges can be encoded into the quantum circuit to realize the subsequent quantum computation.

\subsection{QGCN Quantum Circuit Model}

\begin{figure}[!htb]
	\centering
	\includegraphics[width=0.7\linewidth]{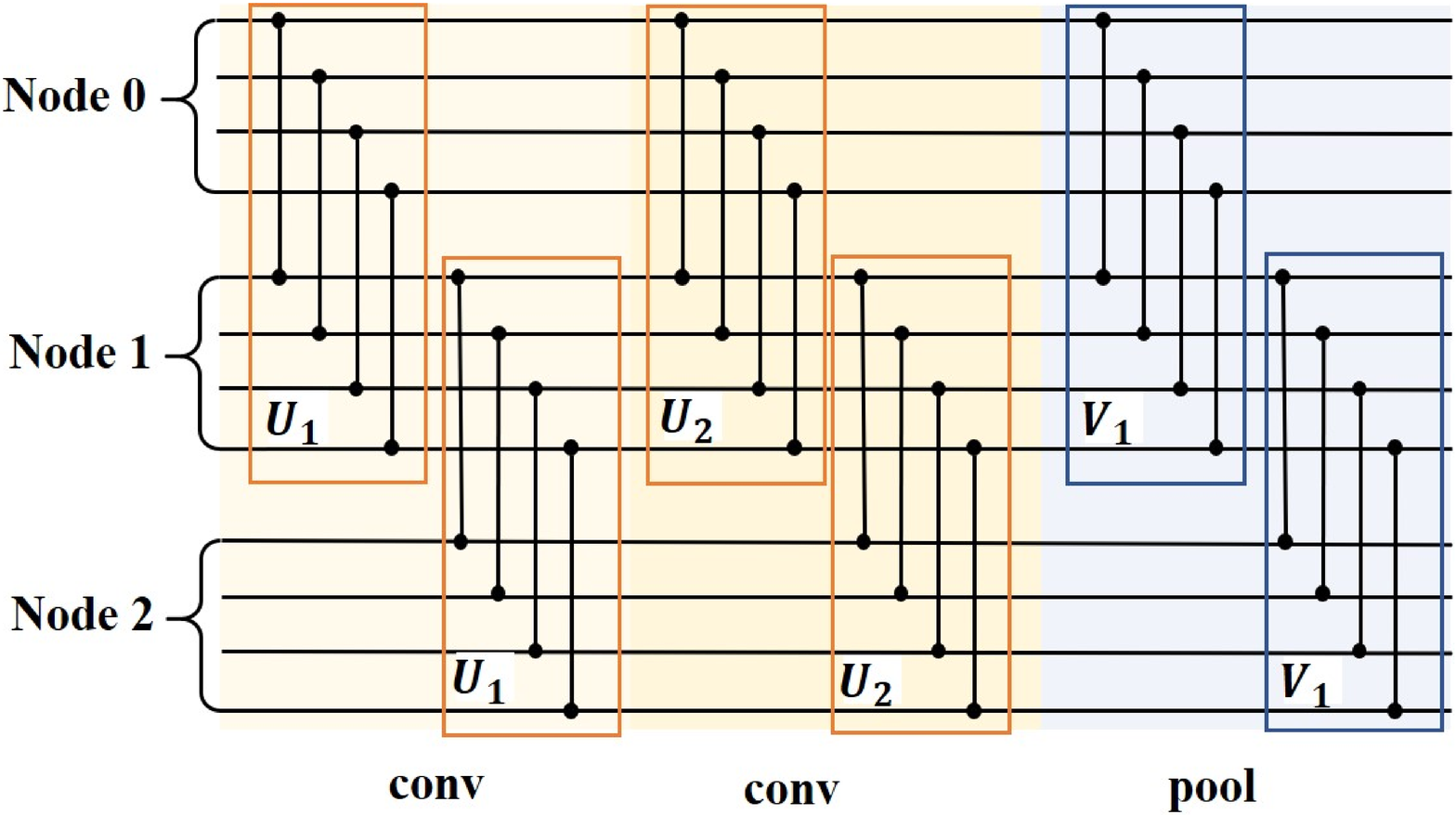}
	\caption{The implementation of QGCN quantum circuit. The connected lines indicate that there are $U$ or $V$ operations between the features of the nodes corresponding to the line.}
	\label{fig9}
\end{figure}

The graph data structure is constructed by the MNIST dataset shown in Fig.~\ref{fig8}(b). Each node includes 16-dimensional (4$\times $4) features, corresponding to 4 qubits. Based on the relationship of edges, the controlled unitary gates control which nodes and which corresponding features are convoluted and pooled. For the specific circuit implementation, refer to the convolutional layer and pooling layer design of quantum graph neural networks in Sections 3.2 and 3.3. According to the datasets preprocessing and the quantum state preparation of node features, the $U$ and $V$ operations in the quantum graph convolutional neural network model correspond to the quantum circuit, as shown in Fig.~\ref{fig9}.

The number of the training interactions is 1000, and the batch size is 16. The squared loss and accuracy are used as the evaluation indicators of the graph classification task.

\subsection{Experimental Results and Discussion}

(1)Nodes selection

\begin{figure}[!htb]
	\centering
	\includegraphics[width=\hsize]{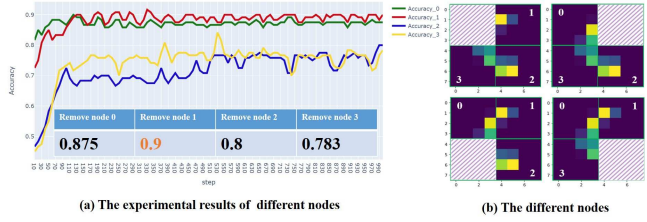}
	\caption{The experimental results of different nodes}
	\label{fig10}
\end{figure}

Three nodes are randomly selected from each sample to construct the graph data structures and analyze which set of nodes has a more critical influence on the classification results through four sets of comparative experiments. The nodes selection method and experimental results are shown in Fig.~\ref{fig10}.

The experimental results show that the selection of different nodes and edges will affect the classification results. The experiment uses the constructed graph structure as the dataset. The model learns the 3/4 features of the entire graph to achieve the classification task, so different features will affect the classification effect. As in Fig.~\ref{fig10}, it is a handwritten picture with label 6. The features of square 1 are less than others, so it has less influence on recognizing the whole picture.

The purpose of the experiment is to test the learning ability of the quantum graph convolutional neural network model. Simple data sets are more comfortable to train. Although the panorama of the image is not used, the relationship between features is more prominent through the location information, which can better reflect the characteristics of graph data structure.

The comparative experiments of node selection indicate that the quantum graph convolutional neural networks strongly depend on the experimental dataset. The expression of the model to the graph structure depends on the network architecture of the graph data itself.

(2) Model complexity analysis

\begin{figure}[!htb]
	\centering
	\includegraphics[width=0.8\hsize]{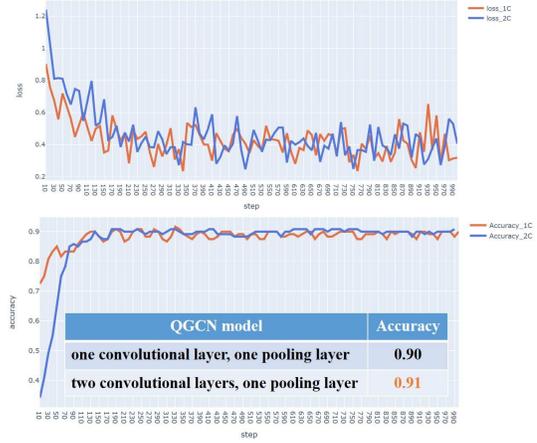}
	\caption{The loss value and accuracy of two models}
	\label{fig11}
\end{figure}

The experiment analyzes the relationship between the number of QGCN network layers and the accuracy of the test set. There are only second-order neighbor nodes at most in the experimental dataset. Therefore, two sets of comparative experiments are designed for the QGCN model: one convolutional layer, one pooling layer, and two convolutional layers, one pooling layer. The experimental results are shown in Fig.~\ref{fig11}.

The experimental results show that the loss values of the two models decrease steadily and converge, indicating that the QGCN model has good learning and network generalization abilities.

Comparing the two models from the accuracy index, one convolution layer, one pooling layer model needs 21 parameters, and two convolution layers, one pooling layer model needs 36 parameters. Increasing the complexity of the model does not significantly improve the accuracy. The possible reason is that the experimental test data is relatively simple. The graph structure only contains three nodes and two edges, and the results of aggregating their first-order neighbors and second-order neighbors are not significantly different.

However, the depth of the graph is deep for complex graph structure, and only aggregating its low-order neighbors can not fully express the graph. Therefore, the 0.1 improvements of the verification experiment can show that the QGCN model can be extended. The model complexity can increase with the depth of the graph, offering better generalization performance. Therefore, QGCN is potential to train deeper networks to achieve more complicated graph classification tasks. However, that would require quantum computers or processors in larger size.

\section{Conclusion and Future Works}

This paper proposes a quantum graph convolutional neural network model to process graph type data, reproducing the classical graph convolutional neural networks by parameterized quantum circuits. It is shown that the structures of four quantum layers can effectively capture node connectivity and learn the hidden layer representation of node features. Numerical simulation results on a graph dataset demonstrate that the proposed model can be effectively trained and has good performance in graph level classification tasks.

The existing graph convolution formulas only use the information of nodes without considering the features of edges. Considering extending quantum graph convolutional neural networks to learn the features of nodes and edges simultaneously will more completely express the graph structure  and will be our future research focus.

\balance


\begin{thebibliography}{0}
\bibitem{Biamonte2017}
J.~Biamonte, P.~Wittek, and N.~Pancotti, Quantum machine learning, \emph{Nature}, 549(7671): 195-202, 2017.

\bibitem{wurebing2017} S.~Lu, Y.~Zheng, X.~Wang, and R.~Wu, Quantum machine learning (in Chinese), \emph{Control Theory \& Applications}, 34(11): 1429-1436, 2017.

\bibitem{Gao2020} Y.~L\"u, Q.~Gao, J.~L\"u, Y.~Pan, and D.~Dong, Recent advances of quantum neural networks on the near term quantum processor (in Chinese), \emph{Scientia Sinica Technologica}, doi: 10.1360/SST-2020-0459, accepted.

\bibitem{Kak1995}
S.~C.~Kak, On Quantum neural computing, \emph{Information Science}, 83: 143-160, 1995.

\bibitem{wurebing2020} R. Wu, X. Cao, P. Xie, and Y. Liu, End-to-end quantum machine learning implemented with controlled quantum dynamics,  \emph{Physical Review Applied}, 14(6): 064020, 2020.

\bibitem{Kapoor2016}
A.~Kapoor, N.~Wiebe, and K.~Svore, Quantum perceptron models, in \emph{Proceedings of the 30th International Conference on Neural Information Processing Systems}, 2016: 4006-4014.

\bibitem{Schütt2017}
K.~T.~Schütt, F.~Arbabzadah, and S.~Chmiela, Quantum--chemical insights from deep tensor neural networks, \emph{Nature communications}, 8(1): 1-8, 2017.

\bibitem{Cong2019}
I.~Cong, S.~Choi, and M.~D.~Lukin, Quantum convolutional neural networks, \emph{Nature Physics}, 15(12): 1273-1278, 2019.

\bibitem{Kipf2016}
T.~N.~Kipf and M.~Welling, Semi-supervised classification with graph convolutional networks, \emph{arXiv preprint}, arXiv:1609.02907, 2016.

\bibitem{Bravyi2018}
S. Bravyi, D. Gosset, and R. König, Quantum advantage with shallow circuits, \emph{Science}, 362(6412): 308-311, 2018.

\bibitem{Zhang2020}
K.~Zhang, M.~H.~Hsieh, and L.~Liu, Toward trainability of quantum neural networks, \emph{arXiv preprint}, arXiv:2011.06258, 2020.

\bibitem{Chung2000} M.~A.~Nielsen and I.~L.~Chung, \emph{Quantum Computing and Quantum Information}. New York: Cambridge University Press, 2000, 61-73.

\bibitem{Krizhevsky2017}
A.~Krizhevsky, I.~Sutskever, and G.~E.~Hinton, ImageNet classification with deep convolutional neural networks, \emph{Communications of the ACM}, 60(6): 84-90, 2017. 

\bibitem{Defferrard2016}
M.~Defferrard, X.~Bresson, and P.~Vandergheynst, Convolutional neural networks on graphs with fast localized spectral filtering, \emph{arXiv preprint}, arXiv:1606.09375, 2016.

\bibitem{Killoran2019}
N.~Killoran, T.~R.~Bromley, and J.~M.~Arrazola, Continuous-variable quantum neural networks, \emph{Physical Review Research}, 1(3): 33063, 2019.

\bibitem{Tacchino2019}
F.~Tacchino, C.~Macchiavello, and D.~Gerace, An artificial neuron implemented on an actual quantum processor, \emph{npj Quantum Information}, 5(1): 1-8, 2019.

\bibitem{Park2019}
D. K. Park, F. Petruccione, and J. K. K. Rhee, Circuit-based quantum random access memory for classical data, \emph{Scientific reports}, 9(1): 1-8, 2019.

\bibitem{Shende2004}
V.~V.~Shende, I.~L.~Markov, and S.~S.~Bullock, Minimal universal two-qubit controlled-NOT-based circuits, \emph{Physical Review A}, 69(6): 062321, 2004.

\bibitem{Vidal2004}
G.~Vidal and C.~M.~Dawson, Universal quantum circuit for two-qubit transformations with three controlled-NOT gates, \emph{Physical Review A}, 69(1): 010301, 2004.

\bibitem{Ville2020}
B.~Ville, I.~Josh, and S.~Maria, PennyLane: Automatic differentiation of hybrid quantum--classical computations, \emph{arXiv preprint}, arXiv:1811.04968, 2018.

\bibitem{LeCun1998}
Y.~LeCun, L.~Bottou, and Y.~Bengio, Gradient-based learning applied to document recognition, in \emph{Proceedings of the IEEE}, 1998: 2278-2324.
\end{thebibliography}
\end{document}